\documentclass[reprint,amsmath,amssymb,aps,prl,longbibliography,lengthcheck,]{revtex4-1}

\usepackage{graphicx}
\usepackage{epsfig}

\begin{document}


\title{Beam asymmetry $\Sigma$  measurements on the $\pi^-$ photoproduction off neutrons}


\author{The Graal collaboration: G. Mandaglio$^{1,2}$, F. Mammoliti$^{2,3}$, M. Manganaro$^{1,2}$, V. Bellini$^{2,3}$, J. P. Bocquet$^4$, L. Casano$^5$, A. D'Angelo$^{5,6}$, R. Di Salvo$^5$, A. Fantini$^{5,6}$, D. Franco$^{5,6}$, G. Gervino$^7$, F. Ghio$^8$, G. Giardina$^{1,2}$, B. Girolami$^8$, A. Giusa$^{2,3}$, A. Ignatov$^9$,  A. Lapik$^9$, P. Levi  Sandri$^{10}$, A. Lleres$^4$, D. Moricciani$^5$, A. N. Mushkarenkov$^9$, V. Nedorezov$^9$, C. Randieri$^{2,3}$, D. Rebreyend$^4$, N. V. Rudnev$^9$, G. Russo$^{2,3}$, C. Schaerf\,$^{5,6}$, M. L. Sperduto$^{2,3}$, M. C. Sutera$^2$, A. Turinge$^9$, V. Vegna$^{5,6}$.\\W. J. Briscoe$^{11}$ and I. I. Strakovsky$^{11}$}
\affiliation{$^{1}$Dipartimento di Fisica, Universit\`a di Messina, I-98166 Messina, Italy}
 \affiliation{$^{2}$INFN, Sezione di Catania, I-95123 Catania, Italy}
   \affiliation{$^{3}$Dipartimento di Fisica e Astronomia, Universit\`a di Catania, I-95123 Catania, Italy}
\affiliation{$^{4}$LPSC, Universit\'e Joseph Fourier, CNRS/IN2P3, Institut National Polytechnique de Grenoble, 38026 Grenoble, France}
\affiliation{$^{5}$INFN, Sezione di Roma``Tor Vergata'', I-00133 Roma, Italy}
 \affiliation{$^{6}$Dipartimento di Fisica, Universit\`a di Roma ``Tor Vergata'', I-00133 Roma, Italy}
\affiliation{$^{7}$Dipartimento di Fisica Sperimentale, Universit\`a di Torino, and INFN - Sezione di Torino, I-00125 Torino, Italy}
\affiliation{ $^{8}$Istituto Superiore di Sanit\`a, I-00161 Roma, and  INFN - Sezione di Roma, I-00185 Roma, Italy}
\affiliation{$^{9}$Institute for Nuclear Research, RU-117312 Moscow, Russia}
 \affiliation{$^{10}$INFN-Laboratori Nazionali di Frascati, I-00044 Frascati, Italy}
\affiliation{$^{11}$ Center for Nuclear Studies
Department of Physics
The George Washington University
Washington, DC 20052, USA}

\date{\today}

\begin{abstract}
The $\Sigma$ beam asymmetry in the photoproduction of negative pions from quasi-free neutrons in a deuterium target was measured at Graal in the energy interval 700 - 1500 MeV and a wide angular range, using polarized and tagged photons. The results are compared with  recent partial wave analyses. 
\end{abstract}


\maketitle


During the last twenty years, pseudoscalar meson photoproduction has proven to be a valid and complementary approach to hadronic reactions for the study of properties of baryon resonances. 
The main disadvantage of the e.m. probe, i.e. the lower cross section values, has been overcome thanks to the advent of a new generation of  high duty cycle electron accelerators and to the resulting high intensity real and virtual photon beams. These beams, in combination with large solid angle and/or large momentum acceptance  detectors, provided recently a large amount of high precision data.

Pseudoscalar meson photoproduction can be described in terms of four complex CGLN\cite{Chew57}
(or equivalently helicity) amplitudes, providing seven real independent
quantities for each set of incident photon energy and meson polar angle
in the CM system. 
To resolve the ambiguities in the context of Barker \textit{et al}. \cite{Barker} it is
necessary to perform a 'complete experiment'. That is eight polarization observables (including the unpolarized differential cross section) need
to be measured for each isospin channel.
 Waiting for
such an experiment, the analysis of meson photoproduction has been concentrated on a description of the reaction mechanisms in terms of intermediate states, which have definite parity and angular momentum
and are therefore excited via electric and magnetic multipoles.

Polarization observables, accessible with the use of polarized photon beams and/or nucleon targets and/or the measurement of the polarization of the recoil nucleon,  play a special role in the disentanglement of the hadron resonances contributing to the reaction\cite{Arndt90,Benmerrouche95,Saghai97,Drechsel99,Feuster99}.

One further complication in the study of meson photoproduction on the nucleons comes from the isospin, which must be conserved at the hadronic final vertex, while it can be changed at the photon vertex. 

In particular for isovector mesons, such as pions, the transition operator can be split into an isoscalar ($\Delta I=0$) and isovector ($\Delta I=1$) components, giving rise to three independent matrix elements $<I_f,I_{f,z}|A|I_i,I_{i,z}>$ describing the transitions between  the initial and final states: one isoscalar $A^{IS}$ (with $\Delta I$ and $\Delta I_3 =0$)
 and two isovector $A^{IV}$ and $A^{V3}$ 
($\Delta I=1$ and $\Delta I_3=0$, $\pm 1$) components.

It is necessary to perform experiments on the proton and neutron for each
final state isospin channel in order to disentangle these transition
amplitudes \cite{watson54,walker69}.


Data on the four reactions ($\gamma p\to\pi^0p$, $\gamma p\to\pi^+n$, $\gamma n\to\pi^0n$, $\gamma n\to\pi^-p$)  have been collected  at Graal, with a polarized photon beam impinging on a H$_2$ or  D$_2$ liquid target and with the final products detected in a large solid angle apparatus. This allowed for the first time the simultaneous extraction of the beam asymmetry values of the four reactions with the same experimental conditions and the same photon energy range (0.55-1.5 GeV), corresponding to the second and third nucleon resonance regions. 
Results for the first three reactions have already been published by the Graal collaboration \cite{slava02,Annich05,paolino09}, providing for the pion photoproduction on the nucleon a very extensive database of high precision data which is composed of: 830 differential cross section and 437 beam asymmetry points for $\pi^0$ photoproduction on the free proton; about 300 beam asymmetry points for $\pi^+$n photoproduction on the free proton; 216 asymmetry points for $\pi^0$ photoproduction both on the quasi-free proton and neutron.

The last of the four reactions is the subject of the present article. The extraction of the beam asymmetry values for $\pi^-$p photoproduction on the quasi-free neutron advances the isospin study of pion photoproduction on the nucleon, constraining the determination of the three isoscalar and isovector transition amplitudes.      


\section{Experimental set-up}

The Graal  $\gamma$-ray beam at the ESRF is produced by the backward scattering in flight of laser photons on the relativistic electrons circulating in the storage ring. This technique, first used on a storage ring for the Ladon beam on the Adone at Frascati  \cite{federici}, produces polarized and tagged  $\gamma$-ray beams with very high polarization and good energy resolution. At its maximum energy the beam polarization is very close to the one of the laser photons (linear or circular) \cite{Caloi} and can be easily rotated or changed with conventional optical components changing the polarization of the laser light. It remains above 74\% for photon energy above 70\% of its maximum. With the 6.03 GeV ESRF accelerator and the 351 nm line of an argon (Ion) Laser, the maximum  $\gamma$-ray energy obtainable is 1487 MeV and the spectrum is almost flat over the whole tagged spectrum. The energy resolution of the tagged beam is limited by the optics of the ESRF magnetic lattice and is 16 MeV (FWHM) over the entire spectrum.

The Graal apparatus has been described in several papers  \cite{slava02,Annich05,paolino09,bart07,fanti2008,Lleres09}. A cylindrical liquid hydrogen (or deuterium) target is located on the beam and coaxial with it. The detector covers the entire solid angle and is divided into three parts. The central part, $25^\circ < \theta  \le 155^\circ$, is covered by two cylindrical wire chambers, a Barrel made of 32 plastic scintillators and a BGO crystal ball made of 480 crystals which is well suited for the detection of $\gamma$-rays of energy below 1.5 GeV. The chambers, the Barrel and the BGO are all coaxial with the beam and the target. The wire chambers detect and measure the positions and angles of the charged particles emitted by the target while the scintillating Barrel measures their energy loss. The BGO ball detects charged and neutral particles and measures the energy deposited by them. For neutral particles it provides a measurement of their angles by its granularity (480 crystals: 15 in the  $\theta$ direction and 32 in the $\phi$  direction).

At forward angles, $\theta  \le 25^\circ$, the particles emitted from the target encounter first two plane wire chambers which measure their angles, then, at 3 meters from the center of the target, two planes of plastic scintillators, made of 26 horizontal and 26 vertical bars to measure the particles position, specific ionization and time of flight, and then a thick (Shower Wall) wall made of a sandwich of scintillators and lead to detect charged particles, $\gamma$-rays and neutrons. The TOF resolution of these scintillators is of the order of 560 ps (FWHM) for charged particles and 900 ps for neutrons. The total thickness of the plastic scintillators is 20 cm and the detection efficiency is about 20\% for neutrons and 95\% for $\gamma$-rays.

Backward angles, $\theta  > 155^\circ$, are covered by two disks of plastic scintillators separated by 6 mm of lead to detect charged particles and gamma-rays escaping in the backward direction. 

The energy of the $\gamma$-rays is provided by the tagging set-up which is located inside the ESRF shielding, attached to the ESRF vacuum system. The electrons which have scattered off a laser photon and produced a $\gamma$-ray have lost a significant fraction of their energy and therefore drift away from the equilibrium orbit of the stored electrons and finally hit the vacuum chamber of the storage ring. Before hitting the vacuum chamber they are detected by the tagging system, which measures their displacement from the equilibrium orbit. This displacement is a measure of the difference between their energy and that of the stored electron beam and therefore provides a measure of the energy of the gamma-ray  produced. 
The tagging system \cite{Annich05} consists of 10 plastic scintillators and a 128 channels Solid State Microstrip Detector with a pitch of  300  $\mu$m. The plastic scintillators signals are synchronized by GaAs electronics  with the RF accelerating system, and provide a timing for the entire electronics of the Graal apparatus with a resolution of 180 ps (FWHM). This allows clear discrimination between electrons coming from two adjacent electron bunches that are separated by 2.8 ns. The Microstrips provide the position of the scattered electron and therefore the energy of the associated gamma-ray. Their pitch (300 $\mu$m) has been set in order to limit the number of tagging channels without appreciable reduction of the gamma-ray energy resolution imposed by the characteristics of the storage ring. The detector is located inside a shielding box positioned in a modified section of the ring vacuum chamber. The shielding box is positioned at 10 mm from the circulating electron beam. This limits the lowest tagged gamma-ray energy to about 550 MeV.

\section{Event identification and data analysis}
\label{mea}

The data analysis is based on the following direct measurements: the energy, $E_{\gamma}$, of the incident photon measured
by the tagging detector; the energy, $E_p$, of the proton measured in the BGO
or by the TOF in the forward wall; the polar and azimuthal angles $\theta_p$
and $\phi_p$ of the proton and $\theta_{\pi^-}$ and $\phi_{\pi^-}$ of the pion measured by the planar and cylindrical MWPCs \cite{mandy09}. The energy of the pion, $E_{\pi^-}$, is obtained by the reaction energy balance  neglecting the Fermi energy of the neutron in the deuterium target ($E_{\pi^-}=E_{\gamma}+M_n-E_p$).

 The charged particle identification in the central part of the apparatus ( 25$^\circ < \theta \le $155$^\circ$) was performed using a cut in the bi-dimensional plot of the energy lost in the barrel versus the energy measured by the BGO calorimeter\cite{mandyproc}. In the forward direction ($\theta \le$ 25$^\circ$) it
was obtained using the bi-dimensional cut on  energy lost versus TOF measured by the plastic scintillator wall\cite{mandyproc}. We also applied to each detected charged particle the condition that a coincidence of the signals from the three charged particle detectors is obtained.

 Our simulation, based on GEANT3\cite{GEANT} and on a realistic event generator\cite{Corvi} has shown that, with the preliminary selection of the  events  obtained by the constraint that proton and pion are the only charged particles detected in the Graal apparatus, the number of events coming from other reaction channels is lower than 14\%.\vspace{0.3cm}

\begin{figure}[htb]
\centering
\vspace{-0.5cm}
\resizebox*{6.cm}{!}{\includegraphics{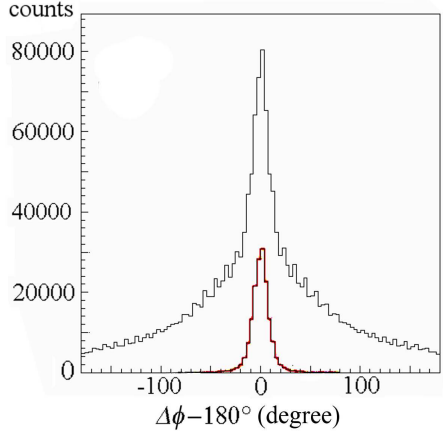}}\vspace{-0.5cm}
\caption{ (Color on line) a) the $\pi^-$-p coplanarity before (upper curve) and after the cuts (lower curve).}\vspace{-0.3cm}
\label{coplan}
\end{figure}

The quantities measured in the Graal-experiment exceed the number required for a full kinematical reconstruction of the event in a (quasi-)two-body kinematics. Therefore it is possible to calculate all kinematic
variables using only a subset of the measured ones. For example,  the polar angle of the pion, $\theta^{calc}_{\pi^-}$, and the energy of the proton, $E_p^{calc}$, have been calculated from the other measured quantities and then compared with the results of their direct observations.

\begin{figure}[htb]
\centering
\vspace{-0.3cm}
\resizebox*{6cm}{!}{\includegraphics{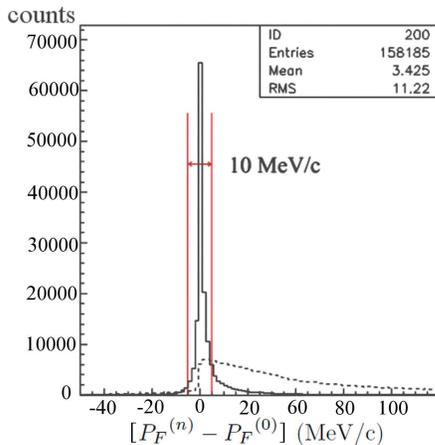}}\vspace{-0.5cm}
\caption{ (Color on line) difference between the Fermi momentum reconstruction at the nth step of
recursive method and at step zero for the signal (solid line) and for the concurrent channels (dashed line) in simulation.}\vspace{-0.3cm}
\label{fermiric}
\end{figure}

Therefore the background from the other reaction channels was drastically reduced with the following constraints:\\
1.\, we  reject all events with additional signals from neutral particles in the BGO or in the Shower Wall;\\
2.\, we impose  coplanarity of the p and $\pi^-$ by the condition $||\phi_{\pi^-}-\phi_p| -180^\circ| < 3 \sigma_\phi$, where $\sigma_\phi$ is the experimental variance of the distribution indicated in  Fig. \ref{coplan};\\
3.\, we impose the condition:
\begin{equation}
\label{threecut}  \sqrt{{\sum_{i}^{x,y,z}{(P_{Fi}-P_{Fi}^{recurs.})}^2}}< 10\; {\rm MeV/c}
\end{equation} 
where: $P_{Fi}$ ($i=x,y,z$) is the component of the Fermi momentum of the target nucleon calculated from the measured kinematical variables neglecting its Fermi energy; $P_{Fi}^{recurs.}$ is its value obtained at the end of a recursive process in which at each stage the Fermi momentum is calculated by inserting into the energy-momentum conservation equations the value of the Fermi energy derived by the value of the Fermi momentum resulting from the previous iteration. The iterations stop when the difference of the modules of the Fermi momentum in two successive iterations is less than 10 keV/c. The cut value, 10 MeV/c, was suggested by the simulation in order to minimize the loss of good events (see Fig. \ref{fermiric});\\
\begin{figure}[htb]
\centering
\vspace{-0.1cm}
\resizebox*{6.cm}{!}{\includegraphics{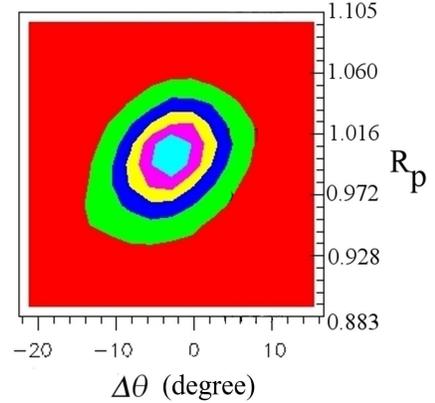}}\vspace{-0.4cm}
\caption{ (Color on line) the bi-dimensional distribution of $\Delta \theta$ vs. $R_p$ as defined in this article.}\vspace{-0.3cm}
\label{thetene}
\end{figure}
4.\, The last constraint is :
\begin{equation}    \frac{{(x-\mu_x)}^2}{\sigma_x^2}+\frac{{(y-\mu_y)}^2}{\sigma_y^2}-
\frac{2C(x-\mu_x)(y-\mu_y)}{\sigma_x \sigma_y} < \sigma^2
\end{equation}
where  $x = \Delta \theta=\theta_{\pi^-}^{calc}-\theta_{\pi^-}^{meas}$; $\theta_{\pi^-}^{meas}$ is the measured angle of the $\pi^-$ while $\theta_{\pi^-}^{calc}$ is the calculated value  from the angle $\theta_{p}$ of the proton and  the gamma-ray energy $E_\gamma$ provided by the tagger; 
 $y=R_p=E_p^{calc}$/$E_p^{meas}$, where $E_p^{meas}$ is the measured value of the proton energy and $E_p^{calc}$ is the calculated value  from $E_\gamma$ and $\theta_{\pi^-}$; $\mu_x$, $\mu_y$, $\sigma_y$ and $\sigma_x$ are the mean values and the variances obtained by a Gaussian fit to the experimental distributions; $C$ is the correlation parameter obtained by a combined best fit of $x$ and $y$ with a bidimensional Gaussian surface(see Figs. \ref{thetene}). 
$\sigma$ has been empirically set at 3, after several attempts, to minimize the loss of good events and at the same time the acceptance of events from competing reactions. As a results the contribution of spurious events is less than 2.3\% as indicated by the simulation. Other systematic errors arise from our imperfect knowledge of the 
beam polarization due to the Laser optics and other minor effects and do not exceed 2\% in total. 

\begin{figure}[htb]
\centering
\vspace{-0.4cm}
\resizebox*{6.cm}{!}{\includegraphics{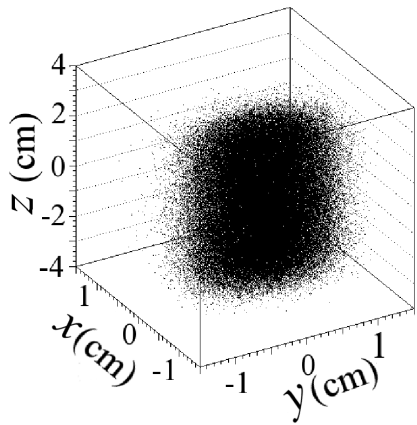}}\vspace{-0.5cm}
\caption{The reconstructed position of the production vertex.}\vspace{-0.3cm}
\label{vert}
\end{figure}

The wire chambers provide the distribution of the reaction vertex\cite{mandy09} inside the deuterium target. Fig.\,\ref{vert} shows that the source of our events is well localized inside the liquid D$_2$.

\begin{figure}[htb]
\centering
\vspace{-0.6cm}
\resizebox*{6cm}{!}{\includegraphics{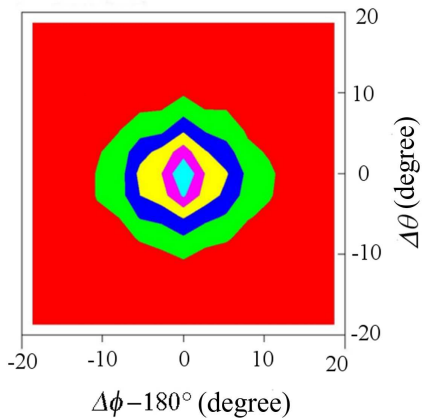}}\vspace{-0.5cm}
\caption{ (Color on line) the bi-dimensional distribution of $\Delta \theta$ vs. $\Delta\phi-180^\circ$=$|\phi_{\pi^-}-\phi_p| -180^\circ$ as defined in this article.}\vspace{-0.3cm}
\label{thetaphi}
\end{figure}

To check the invariance of our results with respect to the selection criteria in a independent analysis we have:\, i) plotted alternatively $\Delta \theta$ vs. $\Delta \phi$-180$^\circ$ which has the advantage that the physical quantities are not correlated as shown in Fig. \ref{thetaphi};\, ii) applied an independent cut  on the variable $R_p$;\, iii) introduced a cut for $P_F\le$ 250 MeV/c instead of the condition 3 - inequality (1). The results of the two procedures are consistent within one standard deviation \cite{manda2010}.

\begin{figure}[htb]
\centering
\vspace{-0.2cm}
\resizebox*{6.5cm}{!}{\includegraphics{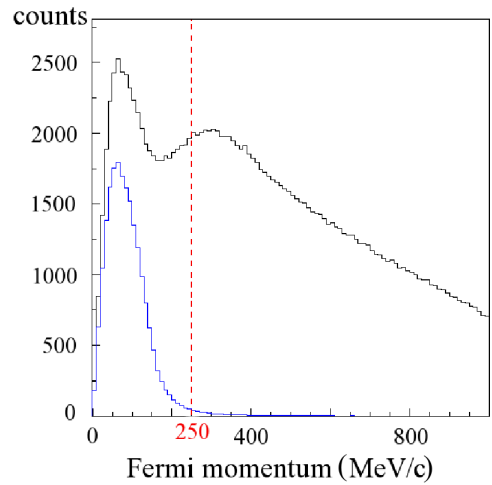}}\vspace{-0.1cm}
\caption{ (Color on line) the Fermi momentum of the neutron calculated before (upper curve) and after (lower curve) the cuts.}\vspace{-0.3cm}
\label{fermicut}
\end{figure}

As a further check we have plotted (see Fig. \ref{fermicut}) the Fermi momentum calculated for all events (spurious included) and that calculated for the ``good'' events (those that have passed our selection). 

The effect of the cuts on the degree of coplanarity of the reaction products and on the Fermi momentum is indicated in Figs. \ref{coplan} and \ref{fermicut} respectively. 

\begin{figure}[htb]
\centering
\resizebox*{6.5cm}{!}{\includegraphics{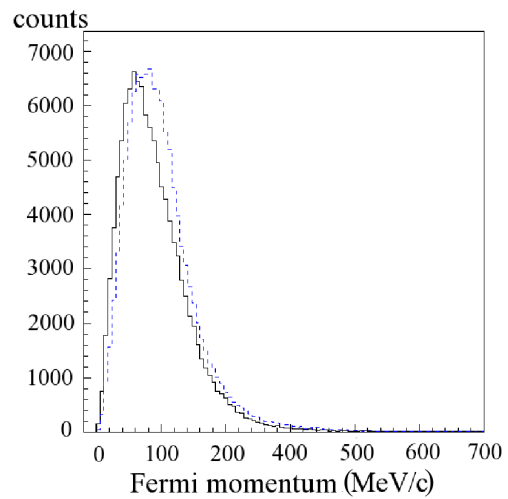}}\vspace{-0.2cm}
\caption{ (Color on line) the Fermi momentum (dashed line) of the neutron after the cuts (data) and that (solid line) generated using the Paris potential.}\vspace{-0.3cm}
\label{fermiparis}
\end{figure}

Our simulation data show that the Gaussian fit of the difference between the Fermi momentum reconstructed by our detectors and the one generated by using the Paris potential  \cite{paris} present a sigma of about 16.9 MeV/c.
The cuts provide a distribution of the Fermi momentum consistent with our knowledge of the structure of the deuteron excluding the spurious events that would require an anomalously large Fermi momentum to satisfy a quasi-two-body kinematic.
Fig.\,\ref{fermiparis} compares the experimental and simulated Fermi momentum distributions. For the simulation we have used the Paris potential \cite{paris} and processed the simulated events through the same analysis software of our data.

The beam polarization asymmetries have been calculated as we did in Refs. \cite{Annich05,fanti2008,paolino09} using the symmetry of the central detector around the beam axis. In the same references are indicated the various checks performed to verify the stability of our results. More than 99\% of the events that survived the cuts produce a reconstructed Fermi momentum lower than 250 MeV/c. For this reason the results presented in Fig. \ref{asym} and in Tab. \ref{tab1} were obtained without any direct cut on the reconstructed Fermi momentum distribution. We refer to \cite{Annich05} for a discussion of possible sources of systematic errors, in particular we have shown that we obtain the same asymmetries, in the overlapping region, using the Green or UV laser lines. These lines produce gamma-ray beams with different spectra and polarizations and their comparison eliminates one main source of systematic errors.     

\begin{widetext}

\begin{table*}[h]
\begin{center}
\resizebox{0.9\columnwidth}{!}{
\begin{tabular}{|r|r|r|r|r|r|r|r|}
\hline
$\theta_{cm} (^{\circ})$ & $E_\gamma$ = 753 MeV & $\theta_{cm} (^{\circ})$ &  $E_\gamma$ = 820 MeV & $\theta_{cm} (^{\circ})$ & $E_\gamma$ = 884 MeV & $\theta_{cm} (^{\circ})$ &  $E_\gamma$ = 947 MeV  \\
\hline
 35.1 &  0.721 $\pm$ 0.053 &  35.0 &  0.504 $\pm$ 0.046 &  35.2 &  0.431 $\pm$ 0.034 &  35.1 &  0.396 $\pm$ 0.026 \\
 52.6 &  0.453 $\pm$ 0.026 &  52.5 &  0.244 $\pm$ 0.015 &  52.5 &  0.219 $\pm$ 0.016 &  52.5 &  0.237 $\pm$ 0.012 \\
 67.2 &  0.267 $\pm$ 0.022 &  66.9 &  0.034 $\pm$ 0.014 &  66.9 & -0.087 $\pm$ 0.017 &  66.7 & -0.074 $\pm$ 0.017 \\
 79.3 &  0.106 $\pm$ 0.019 &  79.1 & -0.173 $\pm$ 0.019 &  79.3 & -0.369 $\pm$ 0.016 &  79.5 & -0.534 $\pm$ 0.020 \\
 89.9 &  0.023 $\pm$ 0.018 &  90.0 & -0.301 $\pm$ 0.019 &  90.3 & -0.426 $\pm$ 0.022 &  90.6 & -0.485 $\pm$ 0.024 \\
104.2 & -0.013 $\pm$ 0.021 & 104.5 & -0.209 $\pm$ 0.023 & 104.6 & -0.218 $\pm$ 0.026 & 105.2 & -0.114 $\pm$ 0.029 \\
127.5 & -0.014 $\pm$ 0.019 & 127.8 & -0.123 $\pm$ 0.021 & 127.3 & -0.189 $\pm$ 0.021 & 127.4 & -0.250 $\pm$ 0.028 \\
148.2 & -0.018 $\pm$ 0.014 & 148.5 & -0.109 $\pm$ 0.010 & 148.7 & -0.174 $\pm$ 0.012 & 149.2 & -0.299 $\pm$ 0.014 \\
162.1 & -0.007 $\pm$ 0.025 & 162.3 & -0.033 $\pm$ 0.029 & 162.4 & -0.080 $\pm$ 0.022 & 162.5 & -0.137 $\pm$ 0.026 \\
\hline
$\theta_{cm} (^{\circ})$ & $E_\gamma$ = 1006 MeV & $\theta_{cm} (^{\circ})$ &  $E_\gamma$ = 1059 MeV & $\theta_{cm} (^{\circ})$ & $E_\gamma$ = 1100 MeV & $\theta_{cm} (^{\circ})$ &  $E_\gamma$ = 1182 MeV  \\
\hline
 35.0 &  0.400 $\pm$ 0.021 &  34.7 &  0.364 $\pm$ 0.016 &  34.3 &  0.329 $\pm$ 0.017 &  33.8 &  0.297 $\pm$ 0.023 \\
 52.5 &  0.226 $\pm$ 0.014 &  52.5 &  0.192 $\pm$ 0.011 &  52.2 &  0.123 $\pm$ 0.011 &  51.1 &  0.083 $\pm$ 0.011 \\
 66.7 & -0.112 $\pm$ 0.014 &  66.7 & -0.185 $\pm$ 0.013 &  66.7 & -0.253 $\pm$ 0.016 &  66.2 & -0.292 $\pm$ 0.011 \\
 79.6 & -0.577 $\pm$ 0.017 &  79.7 & -0.676 $\pm$ 0.024 &  79.8 & -0.636 $\pm$ 0.024 &  79.9 & -0.642 $\pm$ 0.018 \\
 90.7 & -0.447 $\pm$ 0.021 &  90.7 & -0.429 $\pm$ 0.026 &  90.8 & -0.406 $\pm$ 0.027 &  90.7 & -0.388 $\pm$ 0.030 \\
105.1 &  0.094 $\pm$ 0.026 & 104.5 &  0.201 $\pm$ 0.031 & 104.0 &  0.127 $\pm$ 0.039 & 103.2 &  0.149 $\pm$ 0.054 \\
127.5 & -0.248 $\pm$ 0.021 & 127.8 & -0.201 $\pm$ 0.025 & 127.7 & -0.133 $\pm$ 0.034 & 128.1 &  0.032 $\pm$ 0.039 \\
149.4 & -0.374 $\pm$ 0.012 & 149.5 & -0.336 $\pm$ 0.011 & 149.6 & -0.204 $\pm$ 0.014 & 149.6 &  0.079 $\pm$ 0.017 \\
162.6 & -0.113 $\pm$ 0.018 & 162.7 & -0.141 $\pm$ 0.024 & 162.9 & -0.088 $\pm$ 0.026 & 163.0 &  0.131 $\pm$ 0.031 \\
\hline
$\theta_{cm} (^{\circ})$ & $E_\gamma$ = 1259 MeV & $\theta_{cm} (^{\circ})$ &  $E_\gamma$ = 1351 MeV & $\theta_{cm} (^{\circ})$ & $E_\gamma$ = 1438 MeV  \\
\cline{1-6}
 33.6 &  0.259 $\pm$ 0.013 &  33.2 &  0.253 $\pm$ 0.012 &  33.0 &  0.243 $\pm$ 0.011  \\
 50.3 &  0.073 $\pm$ 0.013 &  49.0 &  0.065 $\pm$ 0.008 &  47.6 &  0.045 $\pm$ 0.009  \\
 66.5 & -0.280 $\pm$ 0.017 &  66.6 & -0.236 $\pm$ 0.017 &  66.3 & -0.152 $\pm$ 0.014  \\
 80.0 & -0.493 $\pm$ 0.021 &  79.5 & -0.318 $\pm$ 0.027 &  78.9 & -0.156 $\pm$ 0.023  \\
 90.0 & -0.255 $\pm$ 0.030 &  88.6 & -0.074 $\pm$ 0.026 &  88.5 &  0.035 $\pm$ 0.036  \\
102.1 & -0.042 $\pm$ 0.064 & 102.3 &  0.024 $\pm$ 0.067 & 105.6 &  0.297 $\pm$ 0.100  \\
128.4 &  0.238 $\pm$ 0.040 & 128.5 &  0.529 $\pm$ 0.030 & 127.5 &  0.628 $\pm$ 0.055  \\
149.4 &  0.310 $\pm$ 0.017 & 149.5 &  0.476 $\pm$ 0.021 & 149.3 &  0.489 $\pm$ 0.021  \\
163.1 &  0.229 $\pm$ 0.039 & 163.4 &  0.214 $\pm$ 0.026 & 163.4 &  0.264 $\pm$ 0.029  \\
\cline{1-6}
\end{tabular}
}
\end{center}
\caption{Beam asymmetry $\Sigma$ values for photon energies $E_\gamma$
 ranging from 753 MeV to 1438 MeV. The errors are statistical only.}
\label{tab1}
\end{table*}

\end{widetext}

\section{Results and discussion}
In this paper, we report on the first tagged measurement of the
$\gamma n\to\pi^-p$ reaction by the Graal collaboration in the energy range 
 from 753~MeV to 1438~MeV.  The available 
statistics allowed the determination of the angular sigma-beam asymmetry for 11 bins in the incident-photon energy and 9 angular bins.

\begin{widetext} 
 \begin{figure*}
\vspace{-1 cm}\hspace*{-1.5cm}
 \resizebox*{21cm}{!}{\includegraphics{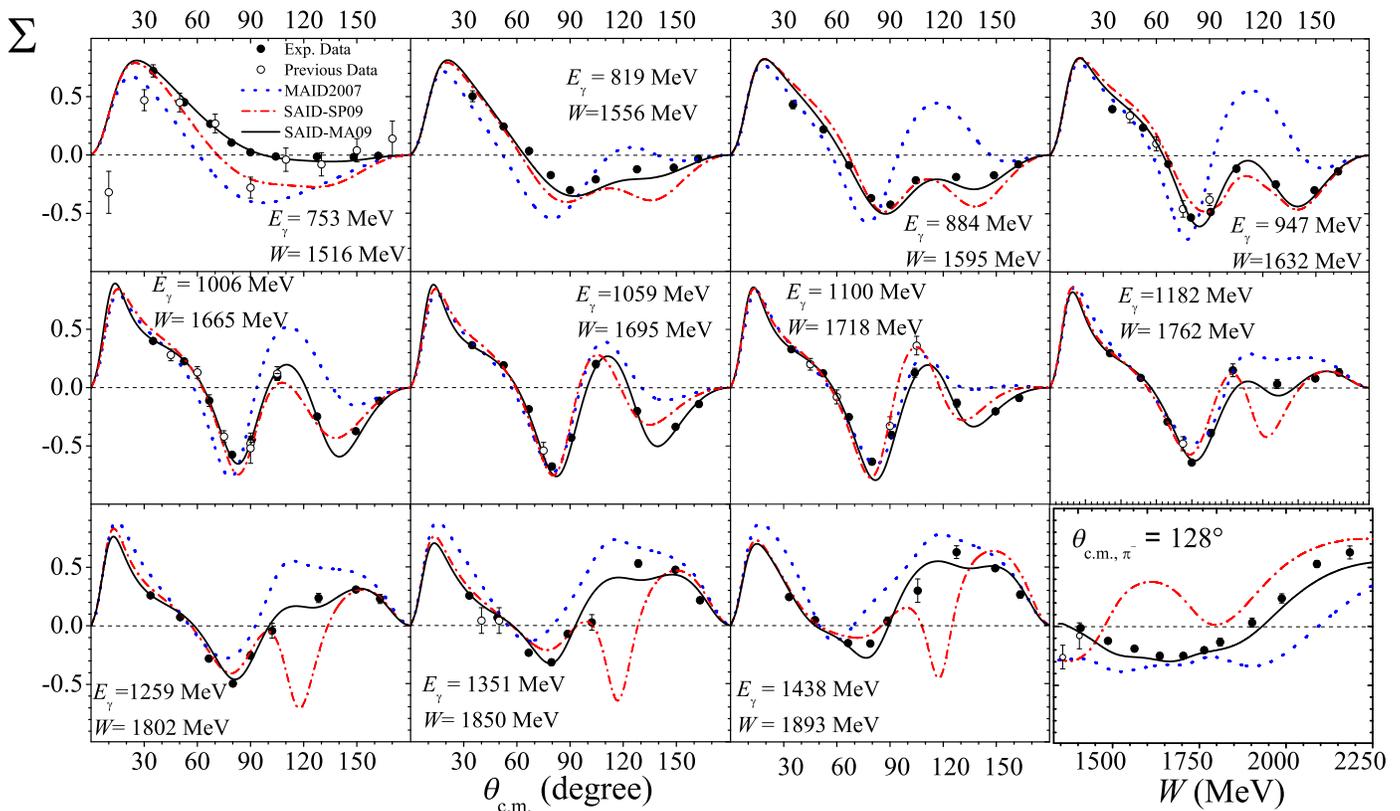}}
\vspace{-1.7cm}
 \caption{\label{asym}(Color on line) The beam polarization asymmetries for $\gamma n\to\pi^-p$ versus pion center-of-mass scattering angle.  The photon energy is shown.  Solid (dash-dotted) lines correspond to the SAID MA09 (SP09 \protect\cite{said}) solution. Dotted lines give the MAID2007 \protect\cite{maid} predictions. Experimental data are from the current (filled circles) and previous measurements\protect\cite{data,yerevan} (open circles).  The plotted points from previously published experimental data are those data points within 4~MeV of the photon energy indicated on each panel. Plotted uncertainties are statistical. In the last panel the asymmetries are plotted versus gamma-ray energy for the CM angle of 128$^\circ$. The MA09 includes in its database the Graal asymmetries for the $\gamma n\to\pi^-p$ and $\gamma n\to\pi^0n$\protect\cite{paolino09} reactions.  SP09 and MAID2007 do not include these data.}\vspace{-0.3cm}
 \end{figure*}
\end{widetext}

Our results for the asymmetries are shown in Fig.~\ref{asym} together with 
previous results~\cite{yerevan} and some theoretical models. The Graal data 
and the results from previous untagged measurements~\cite{yerevan} appear 
to agree well in the overlapping energies. As we have shown in Ref. \cite{paolino09} but also in this analysis we have obtained the same asymmetries using independently two different set of criteria for the event selection.  Moreover the close similarity between the asymmetries measured on the free proton (in hydrogen) and those of quasi-free proton (in deuterium) encourages the assumption that the asymmetries measured on the quasi-free neutron (in deuterium) could be close to those on free neutrons.

Multipole amplitude analyses provide a powerful tool for extracting information 
about the reaction process in as nearly a model-independent manner as possible
\cite{said}.  This approach, in turn, facilitates the identification of
s-channel resonances involved in the reaction process.

SAID-MA09 is the solution that includes our results and recent Graal results 
for $\gamma n\to\pi^0n$~\cite{paolino09} in the best fit while SAID-SP09 does 
not include both of them~\cite{said}. The earlier MAID2007 solution~\cite{maid} 
is also included in Fig.~\ref{asym} for comparison. The status of the 
MAID database for the MAID2007 solution is the same as for SAID-SP09.  
The overall $\chi^2$/Graal data is 483, 2634, and 8793 per 99 Graal 
$\Sigma$s for SAID-MA09, SAID-SP09, and MAID2007 solutions, respectively.  

The SAID-SP09 solution is consistent with our data in the forward angular 
region where previous results constrained the fit. In the backward region 
and at energies above 1100~MeV, the agreement becomes satisfactory only 
after inclusion of our data. The MAID2007 solution agrees with our data 
in the forward region.  Both SAID-SP09 and MAID2007 results exhibit 
structures not seen in the data and 
which explain the poor $\chi^2$ for both cases.

Neutron multipoles from the SAID-MA09 fit are compared to the earlier 
SAID-SP09 determinations in Fig.~3.  Both MA09 and SP09 are quite similar, 
but significant differences between them in magnitude (e.g., 
S$_{11}$, D$_{13}$, and F$_{15}$) are seen.  With the addition of Graal 
$\pi^-$p and $\pi^0$n asymmetries, the SAID solution is now far more 
reliable than in previously published analyses. 

Extending our knowledge of the asymmetry to the backward directions, the results of this experiment constrain the models in the angular region where they had the largest variations and the major differences among themselves.

\begin{widetext}
 \begin{figure*}
\resizebox*{17.5cm}{!}{\includegraphics{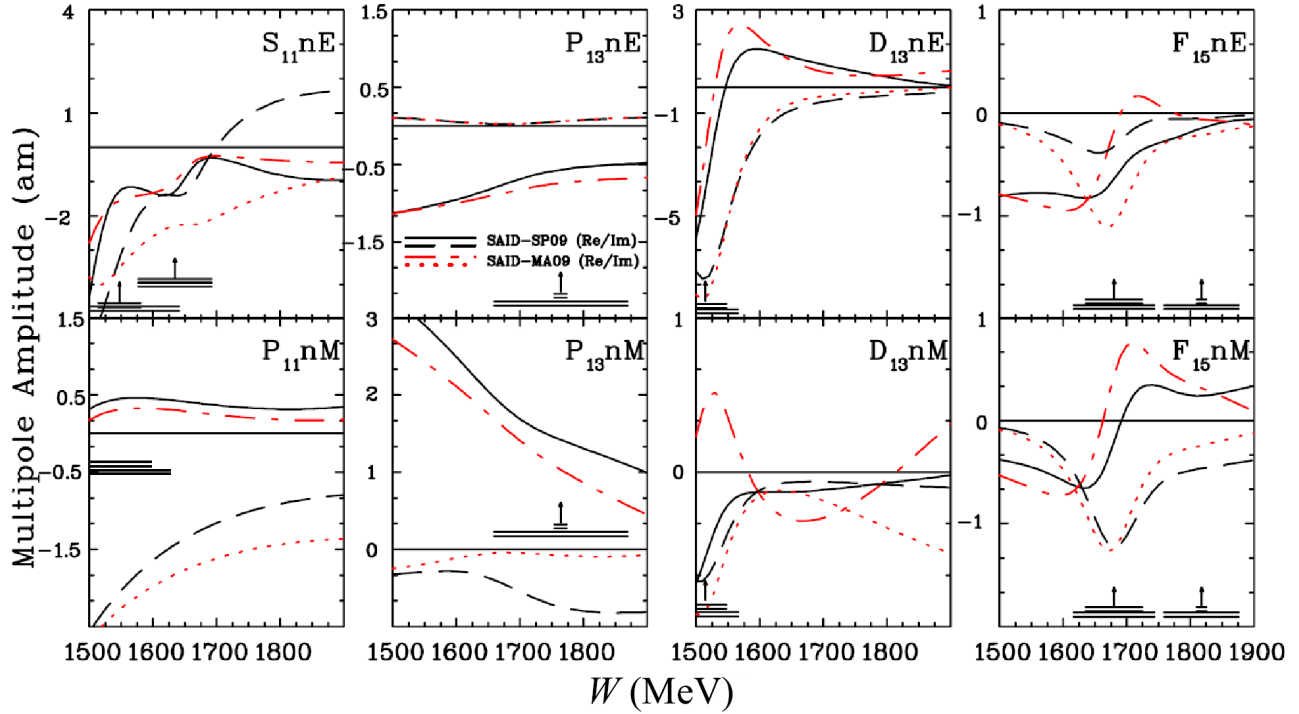}}
 \caption{\label{multip}(Color on line) Multipole amplitudes from $W$ = 1500 to 1900 MeV for isospin 1/2.
Solid (dashed) lines correspond to the real (imaginary) part of
the MA09 solution.  Dashed-dot (dotted) lines give real (imaginary)
part of the SP09~\protect\cite{said} solution.  Vertical arrows
indicate the position of the considered resonance while the horizontal bars show full $\Gamma$ and partial
widths for $\Gamma_{\pi N}$ associated with the SAID $\pi$N solution
SP06~\protect\cite{sp06}.}\vspace{-0.5cm}
 \end{figure*}
\end{widetext}

\begin{acknowledgments}
 We are grateful to the ESRF as a host
institution for its hospitality and the accelerator 
group for the stable and reliable operation of the ring. 
We are very grateful to G. Nobili for his competent and devoted support 
with the realization and maintenance of the experimental apparatus. 
For their technical support we thank:  M. Iannilli, D. Pecchi, E. Tusi and G. Vitali.
WJB and IIS are supported in part by the USDOE DE-FG02-99ER41110 Grant.

\end{acknowledgments}


\end{document}